\documentclass{article}
\pdfoutput=1
\pdfoutput=1 
\usepackage[utf8]{inputenc}
\usepackage{caption}
\usepackage{subcaption}
\usepackage{multicol}
\usepackage{xcolor}
\usepackage{fancyhdr}
\usepackage{cite}
\usepackage{graphics}
\usepackage{wrapfig}
\usepackage{float}
\usepackage{hyperref}
\usepackage{amsmath}
\hypersetup{
    colorlinks,
    citecolor=blue,
    filecolor=blue,
    linkcolor=blue,
    urlcolor=blue
}

\usepackage[bottom]{footmisc}
\usepackage{titling}
\predate{}
\postdate{}

\usepackage{lipsum}
\usepackage{graphicx}

\newcommand{\sqrtsNN}{\mbox{$\sqrt{\mathrm{s}_{_{\mathrm{NN}}}}$}}

\newcommand{\ppt}{$p_{\rm T}$}
\newcommand{\muB}{$\mu_{\rm B}$}

\def \auau  {Au+Au}

\date{} 
\begin{document}


\begin{center}
{\bf Study of strange quark density fluctuations in {\auau} Collisions at {\sqrtsNN} = 7.7 - 200 GeV from AMPT Model }

\vskip1.0cm
Junaid~Tariq$^{1}$ {\footnote{email: junaid.tariq@cern.ch;}},
Sumaira~Ikram$^{2}$ {\footnote{email: sumaira\_ikram90@yahoo.com;}} and
M.~U.~Ashraf$^{3}${\footnote {email: usman.ashraf@cern.ch (Corresponding Author)} }

{\small\it

$^1$ Department of Physics, Quaid-i-Azam University, Islamabad 44000, Pakistan\\
$^2$ Department of Physics, Riphah International University, Islamabad 44000, Pakistan\\
$^3$ Centre for Cosmology, Particle Physics and Phenomenology (CP3), Université Catholique de Louvain, B-1348 Louvain-la-Neuve, Belgium\\

}


\begin{abstract}

The strangeness production is an important observable to study the QCD phase diagram. The yield ratios of strange quark can be helpful to search for the QCD critical end point (CEP) and/or first-order phase transition. In this work, we studied the production of $K^{\pm}$, $\Xi^-(\bar{\Xi}^{+})$, $\phi$ and $\Lambda (\bar \Lambda)$ in {\auau} collisions at {\sqrtsNN} = 7.7, 11.5, 14.5, 19.6, 27, 39, 54.4, 62.4, and 200 GeV from A Multi-Phase Transport model with string melting version (AMPT-SM). We calculated the invariant yield of these strange hadrons using a different set of parameters compared to those reported in earlier studies and also by varying the hadronic cascade time ($t_{max}$) in the AMPT-SM model. We also calculated the yield ratios, $\mathcal{O}_{K^{\pm}-\Xi^{-}(\bar \Xi^{+})-\phi-\Lambda (\bar \Lambda)}$ which are reported as sensitive to the strange quark density fluctuations and found that the AMPT-SM model fails to describe the non-monotonic trend observed by the experimental data. The negative particle ratio are found to be higher than the ratio of positive particles which is consistent with the experimental data. A significant effect is also seen on these ratios by varying the $t_{max}$. For a crossover transition between the Quark-Gluon Plasma (QGP) and hadronic matter, the double yield ratios considered in the present study based on AMPT-SM model do not show any non-monotonic behaviors and thus providing a baseline for the search of CEP, because there is no first-order or second-order phase transition in the AMPT model. The more realistic equation of state based dynamical modeling is still required for the heavy-ion collisions in order to extract the definite physics conclusion about the non-monotonic energy dependence behavior.


\vskip0.5cm

\end{abstract}
\end{center}

\begin{multicols}{2}
\section{Introduction}\label{sec1}

In the recent past years, mapping and understanding the Quantum Chromodynamics (QCD) phase diagram has gained fundamental importance in nuclear physics. Various heavy-ion experiments around the globe are trying to explore the unexplored region of the QCD phase diagram~\cite{BRAHMS:2004adc, PHENIX:2004vcz, PHOBOS:2004zne, STAR:2005gfr, Cleymans:1999st, Becattini:2005xt, Andronic:2005yp}. The main goals of these experiments are to study the properties of the strongly interacting matter, especially the quark-gluon plasma (QGP), the hadronic matter and transition between these two phases~\cite{Shuryak:2014zxa, Braun-Munzinger:2015hba}. Lattice QCD calculations~\cite{Fodor:2004nz} and other effective models~\cite{Asakawa:1989bq, Stephanov:1998dy, Hatta:2002sj} predicts a smooth cross-over between QGP and hadronic matter at vanishing baryon chemical potential ({\muB}) and high temperature ($T$). A first-order phase transition is also predicted at large {\muB} and lower $T$ which ends at possible critical endpoint (CEP)~\cite{Fodor:2004nz, Gavai:2008zr} that separates these two transitions. Various theoretical suggestions have been proposed to search for the CEP~\cite{Asakawa:2008ti, Asakawa:2000wh, Yu:2018kvh, Sun:2017xrx, Sun:2018jhg, Shao:2019xpj}. RHIC has carried out the Beam Energy Scan (BES) program~\cite{STAR:2009sxc, STAR:2010vob, STAR:2019bjj, Ashraf:2016fjl, Ashraf:2021nkb, STAR:2019vcp} by systematic study of {\auau} collisions at {\sqrtsNN} = 7.7 - 200 GeV to find the CEP and search for the phase boundary. The other future experiments, such as the future Facility for Antiproton and Ion Research (FAIR) and Nuclotron-based Ion Collider Facility (NICA) also planned to find/locate the CEP.

Strange hadrons are considered as excellent tool to study the phase boundary and onset of deconfinement. Because the mass of strange quark has same order of magnitude as the temperature of the QGP phase transition and these are produced abundantly in the QGP~\cite{Rafelski:1982pu}. Later the strange quarks are converted to the strange hadrons after hadronization. The strangeness enhancement in $AA$ collisions in contrast with $pp$ collisions has been suggested as a signature of QGP~\cite{Rafelski:1982pu}. Until now, many experiments reported the extensive study of strangeness production at different accelerator facilities~\cite{ Ahmad:1991nv, Ahmad:1998sg, E-802:1999ewk, E917:2001eko, Albergo:2002tn, E895:2003qcm, NA49:2002pzu, NA57:2004nxc, STAR:2002fhx, PHENIX:2002svd, ALICE:2013cdo, Chen:2018tnh}. The yields of strange hadrons observed in $AA$ collisions are close to the expectations from statistical models~\cite{Andronic:2005yp, Becattini:2000jw, Braun-Munzinger:2001alt, Redlich:2001kb}. The ongoing second phase of the BES program will certainly help to precisely measure the yields of strange hadrons in $AA$ collisions and may lead to better understanding of the strangeness production mechanisms at lower energies and QCD phase diagram. In heavy-ion collisions, the search for the CEP is currently of great interest. The QCD phase diagram can also be mapped by studying the fluctuations of physical observables in relativistic nuclear collisions~\cite{Asakawa:2000wh}. Studies based on the transport model~\cite{Li:2016uvu} and hydrodynamical approach~\cite{Steinheimer:2012gc, Steinheimer:2013xxa} shows the spinodal instabilities during the first-order phase transition between the QGP and hadronic matter and possibly can induce significant fluctuations in the baryon density. Thus, the study of particle yield ratios are helpful to probe the significant density fluctuations in baryon density distributions. The collision energy dependence of quark density fluctuations in heavy-ion collisions is an important observable to search for the CEP and pin down the energy where the quark density fluctuations shows a non-monotonic behavior. 

It is reported in Ref.~\cite{Sun:2017xrx, Sun:2018jhg} that the yield ratio $\mathcal{O}_{p-d-t} = N_p N_{^3H}/N^2_{d}$ shows a non-monotonic energy dependence in the framework of nucleon coalescence model for light nuclei and is sensitive to the neutron density fluctuations $\Delta n = \langle (\delta n)^2 \rangle / \langle n\rangle^2$ at kinetic freeze-out. It is suggested in Ref.~\cite{Ko:2018ekv} that, the yield ratio $\mathcal{O}_{K-\Xi-\phi-\Lambda} = \frac{N (K^+) N (\Xi)}{N (\Lambda) N (\phi) }$ in heavy-ion collisions is a sensitive probe to study the quark density fluctuations. This is due to less scattering of strange hadrons compared to the nucleons during the hadronic expansion, and their final abundance at kinetic freeze-out stage are expected to be similar to those at hadronization, if the contribution from the resonance decays are included. A possible strange quark density fluctuations in relativistic heavy-ion collisions can be seen from the non-monotonic energy dependence of this yield ratio.

In this study, we report the energy dependence of the yield ratio $\mathcal{O}_{K^{\pm}-\Xi^{-}(\bar \Xi^{+})-\phi-\Lambda (\bar \Lambda)} $ in {\auau} collisions at {\sqrtsNN} = 7.7, 11.5, 14.5, 19.6, 27, 39, 54.4, 62.4, and 200 GeV from A Multi-Phase Transport model with string melting version (AMPT-SM). The paper is organized as follows: Section~\ref{sec2} gives a brief introduction of the AMPT model. The relation between strange quark density fluctuation and the yield ratio is given in Sec.~\ref{sec3}. Results are presented in Sec.~\ref{sec4} followed by the conclusion in Sec.~\ref{sec5}.


\section{AMPT model}{\label{sec2}}

Relativistic heavy-ion collisions are very complicated and a lot of dynamical evolution is involved from very first stage of the collision to the final state interaction. These evolutionary processes can be explored by using various microscopic transport models by simulation the relativistic heavy-ion collisions. These models include RQMD~\cite{Sorge:1995dp, Sorge:1997nv}, UrQMD~\cite{Bass:1998ca, Bleicher:1999xi}, ARC~\cite{Kahana:1996ssp}, ART~\cite{Li:1997pu} and JAM~\cite{Nara:2019crj}. A Multi-Phase Transport (AMPT) model~\cite{Lin:2004en} is a hybrid transport model~\cite{Lin:2001zk, Chen:2004vha} based on four components: the initial conditions are described by the Heavy-Ion Jet Interaction Generator (HIJING)~\cite{Wang:1991hta}. The parton transport after initialization is described by the Zhang's Parton Cascade (ZPC) model. The quark coalescence model or Lund string fragmentation model~\cite{Andersson:1983jt, Andersson:1983ia} is responsible for the hadronization process in which nearest partons combine together to form hadrons. The last component of the AMPT model is hadronic interactions in which the hadronic re-scatterings are described by a hadronic cascade and is based on A Relativistic Transport (ART) model. Currently, two versions, the default AMPT and AMPT with string melting (AMPT-SM) version are available for studies. The default version of the AMPT was first released around April 2004, whereas the AMPT-SM version was introduced later. More details can be found in Refs.~\cite{Zhang:1999bd, Lin:2000cx, Lin:2021mdn}.\\

In this study we focus on the yield ratios $\mathcal{O}_{K^{\pm}-\Xi^{-}(\bar \Xi^{+})-\phi-\Lambda (\bar \Lambda)}$ in {\auau} collisions at {\sqrtsNN} = 7.7 - 200 GeV from the AMPT-SM version with a new set of parameters~\cite{Ashraf:2022rjt}. The energy dependence of yield ratios of negative particles is also studied. We set the strong coupling constant ($\alpha_s$) to be 0.33 and the parton screening mass ($\mu$) to be 3.20 $f$m$^{-1}$ which led to the value of $\sigma_p$ = 1.5 $mb$. An improved quark coalescence method~\cite{He:2017tla} and ART model for the dynamics of hadronic matter is used to calculate the $\mathcal{O}_{K^{\pm}-\Xi^{-}(\bar \Xi^{+})-\phi-\Lambda (\bar \Lambda)} $ ratios. The potentials of hadrons in the ART model is described by the mean field which is not taken into account to carry the current study~\cite{Xu:2012gf}. In the improved quark coalescence method, a new coalescence parameter $r_{BM}$, which controls the relative probability of a quark forming a baryon rather than a meson is set to 0.61. The baryon either can be produced in a combination with meson ($BM\bar B$ ) or in pairs ($B\bar B$). The baryon production is explained by the popcorn method and, in AMPT-SM model a popcorn parameter PARJ(5) controls the baryon production. The relative of $B\bar B$ and $BM\bar B$ channels is constrained by changing the PARJ(5) from 1.0 (default) to 0.0. Furthermore, the hadronic cascade time ($\tau_{HC}$) and in AMPT-SM model it is referred to as $t_{max}$ also varied from 30 $f$m/$c$ to 0.4 $f$m/$c$ to study its effect on these ratios. The effect of hadron cascade time on particle production mechanisms in {\auau} collisions at {\sqrtsNN} = 7.7 - 200 GeV and $Xe+Xe$ collisions at {\sqrtsNN} = 5.44 TeV is reported in Refs.~\cite{Pradhan:2021zbt, Ashraf:2022rjt}.

\section{Relation between strange quark density fluctuation and yield ratio of strange hadrons}\label{sec3}

In this section, based on Refs.~\cite{Sun:2017ooe, Shao:2019xpj}, we will briefly discussed the relation between strange quark density fluctuations $\Delta s = \langle (\delta s)^2 \rangle / \langle s\rangle^2$ and yield ratios $\mathcal{O}_{K-\Xi-\phi-\Lambda}$ of strange hadrons.  

The hadron yield from the QGP of volume $V_C$ at the phase transition temperature $T_C$ can be calculated by the analytical coalescence formula COAL-SH as reported in Refs.~\cite{Sun:2017ooe, Shao:2019xpj}. According to the constituent quark model, the constituent quarks of the strange hadrons considered to calculate the ratio $\mathcal{O}_{K-\Xi-\phi-\Lambda}$ in the current study are all in s-state ($l = 0$), which leads to $G (l, x) = 1$, where $G (l, x) = \sum_{k=0}^{l} \frac{l!}{l! (l-k)!} \frac{1}{(2k+1) x^{2k}}$ is known as the suppression factor due to coalescence probability and the orbital angular momentum. The yields of considered four strange hadrons can be expressed as: 

\begin{equation}\label{eq1}
    N_{K^+} = g_{K^+} \frac{(m_u + m_{\bar s})^{3/2}}{m_{u}^{3/2} m_{\bar s}^{3/2}} \frac{N_u N_{\bar s}}{V_C} \frac{(2\pi / T_C)^{3/2}}{1+\omega / (2 T_C)}\\    
\end{equation}
\begin{equation}\label{eq2}
    N_{\Xi^-} = g_{\Xi^-} \frac{(m_d + 2m_{s})^{3/2}}{m_{d}^{3/2} m_{s}^{3/2}} \frac{N_d N_{s}}{V_C^2} \frac{(2\pi / T_C)^{3/2}}{[1+\omega / (2 T_C)]^2}\\    
\end{equation}
\begin{equation}\label{eq3}
    N_{\phi} = g_{\phi} \frac{(m_s + m_{\bar s})^{3/2}}{m_{s}^{3/2} m_{\bar s}^{3/2}} \frac{N_s N_{\bar s}}{V_C} \frac{(2\pi / T_C)^{3/2}}{1+\omega / (2 T_C)}\\    
\end{equation}
\begin{equation}\label{eq4}
    N_{\Lambda} = g_{\Lambda} \frac{(m_u + m_d +m_s)^{3/2}}{m_u^{3/2} m_{d}^{3/2} m_s^{3/2}} \frac{N_u N_d N_s}{V_C^2} \frac{(2\pi / T_C)^{3/2}}{[1+\omega / (2 T_C)]^2}\\    
\end{equation}

where, $g_{K^+} = g_{rel}, {K^+}/36$, $g_{\Xi^-} = g_{rel}, {\Xi^-}/108$, $g_{\phi} = g_{rel}, {\phi}/12$ and $g_{\Lambda} = g_{rel}, {\Lambda}/108$. The value of $\omega$ is usually determined by the root-mean-square (RMS) value of the hadron radii~\cite{Shao:2019xpj}. The Eq.(\ref{eq1})-Eq.(\ref{eq4}) can be generalized by introducing the quark density fluctuations, i.e., $n_{q} (\Vec{r}) = \frac{1}{V_C} \int n_{q} (\Vec{r}) d\Vec{r} + \delta n_{q} (\Vec{r}) = \langle q \rangle \delta q (\Vec{r})$. The average value over the coordinate space is represented by $\langle . \rangle$ and $\delta q (\Vec{r})$ with $\langle \delta q \rangle = 0$ is its deviation from the average value $\langle q \rangle$. By introducing the relative quark density fluctuations, $\Delta q = \langle (\delta q)^2 \rangle / \langle q\rangle^2$ and correlation coefficients related to quark density fluctuations in Eq.(\ref{eq1})-Eq.(\ref{eq4}) can be written as:

\begin{equation}\label{eq5}
    N_{K^+} = g_{K^+} \frac{(2\pi / T_C)^{3/2}}{1+\omega / (2 T_C)} V_C \langle \bar s \rangle \langle u \rangle (1+\alpha_{\bar s u}) \\    
\end{equation}
\begin{equation}\label{eq6}
    N_{\Xi^-} = g_{\Xi^-} \frac{(2\pi / T_C)^{3/2}}{[1+\omega / (2 T_C)]^2} V_C \langle s \rangle^2 \langle d \rangle \times (1+\Delta s +2\alpha_{sd})\\    
\end{equation}
\begin{equation}\label{eq7}
    N_{\phi} = g_{\phi} \frac{(2\pi / T_C)^{3/2}}{1+\omega / (2 T_C)} V_C \langle s \rangle \langle \bar s \rangle (1+\alpha_{s \bar s})\\    
\end{equation}
\begin{equation}\label{eq8}
    N_{\Lambda} = g_{\Lambda} \frac{(2\pi / T_C)^{3/2}}{[1+\omega / (2 T_C)]^2} V_C \langle s \rangle \langle u \rangle \langle d \rangle \times (1+\alpha_{sd} + \alpha_{su}+\alpha_{ud})\\    
\end{equation}
 Eq.(\ref{eq5})-Eq.(\ref{eq8}) lead to the expression for the ratio $\mathcal{O}_{K-\Xi-\phi-\Lambda}$:

\begin{multline}\label{eq9}
    \mathcal{O}_{K-\Xi-\phi-\Lambda} = \frac{1}{3} \frac{g_{rel,K^{+}} g_{rel, \Xi^{-}}}{g_{rel, \phi} g_{rel, \Lambda}} \\ \times \frac{(m_u + m_{\bar{s}})^{3/2} (m_d + 2m_{s})^{3/2}}{(m_s + m_{\bar {s}})^{3/2} (m_u + m_d +m_s)^{3/2} }\\ \times \frac{(1+\alpha_{\bar s u}) (1+\Delta s +2\alpha_{sd})}{(1+\alpha_{s \bar s}) (1+\alpha_{sd} + \alpha_{su}+\alpha_{ud})}
\end{multline}

The yield ratio $ \mathcal{O}_{K-\Xi-\phi-\Lambda} = g$, in the absence of quark density fluctuations. The detailed derivation is reported in Ref.~\cite{Shao:2019xpj}. If we neglect the quark correlation coefficients $\alpha_{q1q2}$ of density fluctuation then we have $ \mathcal{O}_{K-\Xi-\phi-\Lambda} = g(1+\Delta s)$, implies the linear proportional relation of the ratio with the strange quark relative density fluctuation $\Delta s$. One can see that strange quark density fluctuation can be probed by measuring the yield ratio of strange hadrons in heavy-ion collisions. Therefore, production of strange hadrons are useful tool to study the QCD phase transition and/or CEP.

\section{Results and Discussion}\label{sec4}

In this section, we present the transverse momentum ({\ppt}) spectra of $K^+, \Xi^-, \phi$ and $\Lambda$ and the yield ratio, $\mathcal{O}_{K^+-\Xi^{-}-\phi-\Lambda}$ in most central (0-10\%) and peripheral (40-60\%) {\auau} collisions at {\sqrtsNN} = 7.7 - 200 GeV from the AMPT-SM model. Additionally, we also studied the energy dependence of yield ratio of negative strange hadrons $\frac{N (K^-) N (\bar \Xi^+)}{N (\bar \Lambda) N (\phi) }$ at different centrality classes from AMPT-SM model. The set of parameters used for this study is discussed in Sec.~\ref{sec2}. 

\begin{figure*}[!ht]
    \centering
    \includegraphics[width=0.75\textwidth]{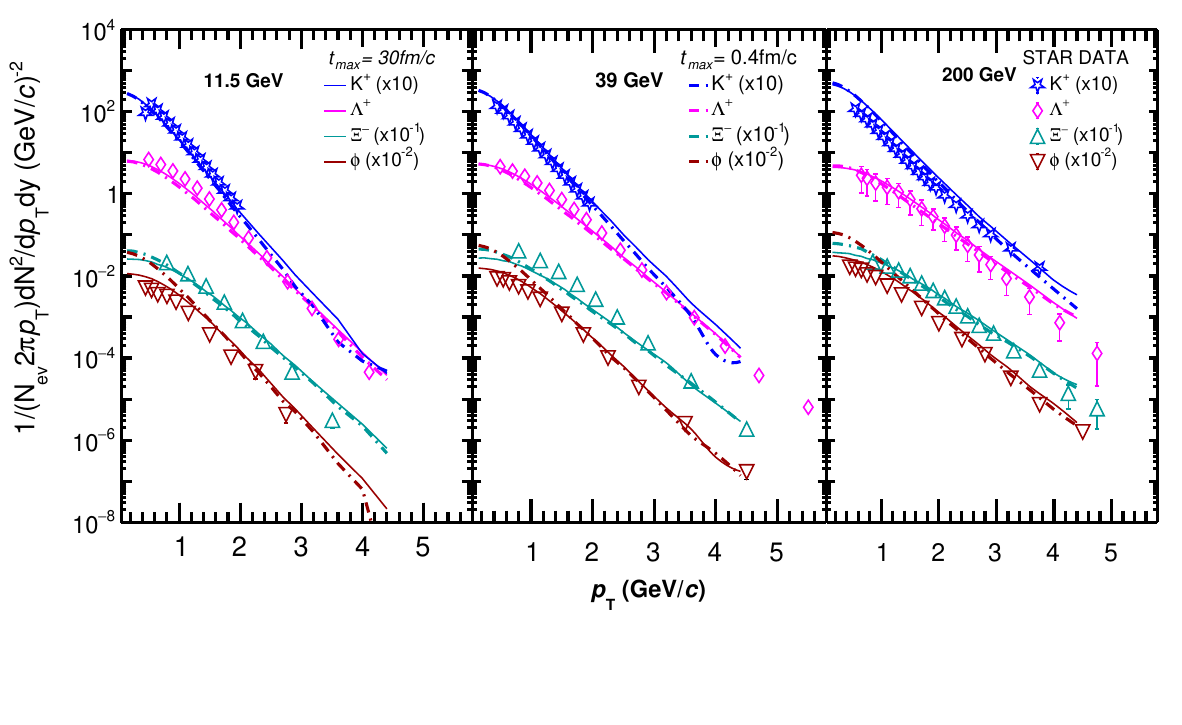}
    \vspace{-3.0\baselineskip}
    \caption{Transverse momentum {\ppt} spectra of $K^+, \Xi^-, \phi$ and $\Lambda$ in most central (0-10\%) {\auau} collisions at {\sqrtsNN} = 11.5, 39 and 200 GeV from the AMPT-SM model for $t_{max} =$ 30 $f$m/$c$ (solid line) and 0.4 $f$m/$c$ (dashed line). The STAR data for comparison is from Refs.~\cite{STAR:2019bjj, STAR:2017sal, STAR:2015vvs, STAR:2006egk, STAR:2007mum, STAR:2002hpr, STAR:2008med}. The {\ppt} spectra is scaled with a constant factor for better visualization. }
    \label{fig1}
\end{figure*}

Figure~\ref{fig1} shows the transverse momentum ({\ppt}) spectra of $K^+, \Xi^-, \phi$ and $\Lambda$ in most central (0-10\%) {\auau} collisions at {\sqrtsNN} = 11.5, 39 and 200 GeV from the AMPT-SM model. The {\ppt} spectra of different particle species are scaled with different constant factor for better visualization. In order to explore the effect of hadronic cascade time ($\tau_{HC}$) on production of these strange hadrons, we varied the $t_{max}$ parameter in the AMPT-SM model from 30 $f$m/$c$ to 0.4 $f$m/$c$. It can be seen that $K^+$ spectra from the experiment is well described by the AMPT-SM model at {\sqrtsNN} = 11.5 and 39 GeV for both values of $t_{max}$, however, the AMPT-SM model slightly over predict the $K^+$ spectra at {\sqrtsNN} = 200 GeV. The $\Lambda$ spectra at {\sqrtsNN} = 11.5 GeV is slightly under predicted by the AMPT-SM model at intermediate {\ppt} bins, and the experimental data at {\sqrtsNN} = 39 and 200 GeV is well described by the AMPT-SM model. The $K^+$ and $\Xi^-$ spectra is not affected significantly by varying the hadron cascade time. The SMPT-SM model under predicts the $\Xi^-$ spectra at {\sqrtsNN} = 11.5 and 39 GeV, particularly at low and intermediate {\ppt}, while a good agreement is seen between the $\Xi^-$ spectra obtained from experimental data and AMPT-SM model at {\sqrtsNN} = 200 GeV. A slight effect of hadronic cascade time $t_{max}$ is also observed in $\Xi^-$ spectra at $p_\mathrm{T} < 1$ GeV/$c$. This effect is more prominent in $\phi$ spectra at very low {\ppt} bins. Overall, there is a good agreement between the experimental data and the AMPT-SM model. This study signifies the importance of various processes involved in the hadronic phase and their impact on the {\ppt} spectra and yield ratios of strange hadrons. This gives us motivation to study the effect of hadronic cascade time on the yield ratio of strange hadrons, $\mathcal{O}_{K^{\pm}-\Xi^{-}(\bar \Xi^{+})-\phi-\Lambda (\bar \Lambda)}$ which is sensitive to the strange quark density fluctuations in heavy-ion collisions.

\begin{figure*}[!ht]
    \centering
    \includegraphics[width=0.75\textwidth]{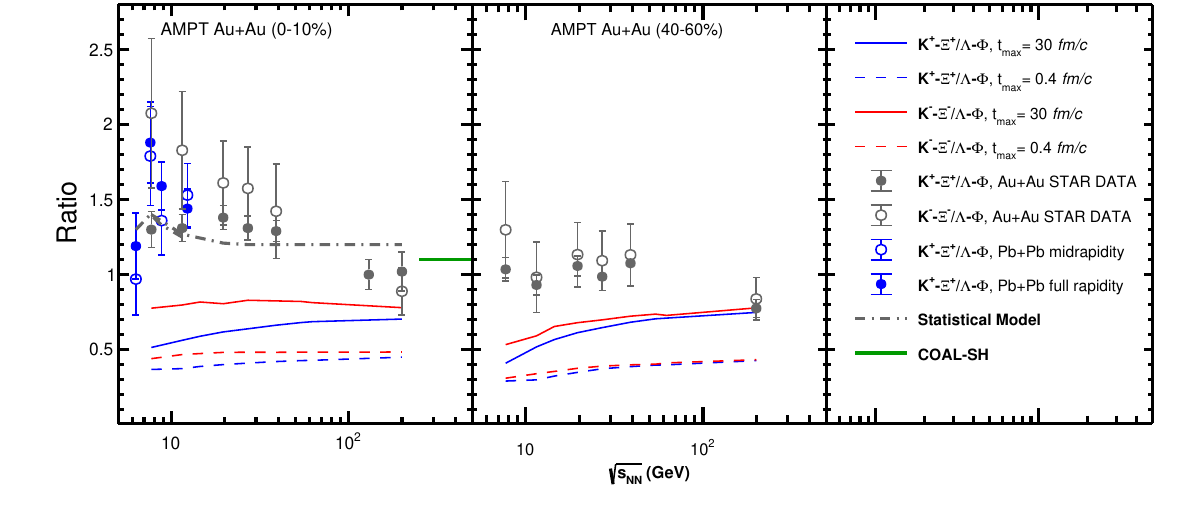}
    \vspace{-1.5\baselineskip}
    \caption{Collision energy ({\sqrtsNN}) dependence of the ratio $\mathcal{O}_{K^{\pm}-\Xi^{-}(\bar \Xi^{+})-\phi-\Lambda (\bar \Lambda)} = \frac{N (K^\pm) N (\Xi^- (\bar \Xi^{+}))}{N (\Lambda (\bar \Lambda)) N (\phi) }$ in most central (0-10\%) and peripheral (40-60\%) {\auau} collisions from the AMPT-SM model for $t_{max} =$ 30 $f$m/$c$ (solid line) and 0.4 $f$m/$c$ (dashed line) at the RHIC energies. The grey dash line for (0-10\%) centrality is the ratio for positive particle calculated from the statistical model~\cite{Wheaton:2004qb}. The green solid horizontal line for (0-10\%) centrality on the right is calculated from the COAL-SH model without quark density fluctuations~\cite{Sun:2017ooe}. The (0-7.2\%) central Pb+Pb collisions data at SPS energies from the NA49 Collaboration are from Refs.~\cite{NA49:2002pzu, NA49:2007stj, NA49:2008goy, NA49:2008ysv}. The STAR data at various energies are from Refs.~\cite{STAR:2019bjj, STAR:2017sal, STAR:2015vvs, STAR:2006egk, STAR:2007mum, STAR:2002hpr, STAR:2008med}.  }
    \label{fig2}
\end{figure*}

Figure~\ref{fig2} shows the energy dependence of yield ratios $\mathcal{O}_{K^{\pm}-\Xi^{-}(\bar \Xi^{+})-\phi-\Lambda (\bar \Lambda)}$ of strange hadrons in most central (0-10\%) and peripheral (40-60\%) {\auau} collisions from the AMPT-SM model for $t_{max} =$ 30 and 0.4 $f$m/$c$ at the RHIC energies. The (0-7.2\%) central Pb+Pb collisions data at SPS energies from the NA49 Collaboration are presented for comparison and are taken from Refs.~\cite{NA49:2002pzu, NA49:2007stj, NA49:2008goy, NA49:2008ysv}. A non-monotonic energy dependence in the yield ratio $\mathcal{O}_{K^+-\Xi^{-}-\phi-\Lambda}$ is clearly seen at {\sqrtsNN} $\approx$ 8 GeV from the experimental data which is consistent with the yield ratio $\mathcal{O}_{p-d-t}$ as reported in Refs.~\cite{Sun:2017xrx, Sun:2018jhg}. This effect becomes insignificant as we go from central to peripheral collisions. The fit to the experimental data with statistical model results $g$= 1.1 from COAL-SH as described in Ref.~\cite{Sun:2018jhg}. The strange quark density fluctuations are not expected at {\sqrtsNN} = 200 GeV and the COAL-SH reasonably describe the experimental data. The statistical model shows a weak energy dependence in the yield ratio $\mathcal{O}_{K^+-\Xi^{-}-\phi-\Lambda}$ in contrast with the experimental data. It is important to point out that the predictions from COAL-SH and statistical model are only for the yield ratio $\mathcal{O}_{K^+-\Xi^{-}-\phi-\Lambda}$ and most central collisions taken from the Ref.~\cite{Sun:2018jhg}. On the other hand, the yield ratios $\mathcal{O}_{K^{\pm}-\Xi^{-}(\bar \Xi^{+})-\phi-\Lambda (\bar \Lambda)}$ calculated from the AMPT-SM model are much smaller than that of the experimental data and shows a weak energy dependence. Similar to the experimental data, the AMPT-SM model also gives higher yield ratio of negative particles as compared to the yield ratio of positive particles. The effect of hadronic casdade time on the yield ratio $\mathcal{O}_{K^{\pm}-\Xi^{-}(\bar \Xi^{+})-\phi-\Lambda (\bar \Lambda)}$ is also studied by varying the $t_{max}$ parameter in the AMPT-SM model from 30 to 0.4 $fm/c$. Both, the positive and negative yield ratios seems to converge towards higher energies for both values of $t_{max}$, however, this effect is less significant at lower energies due to the possible change in underlying particle production mechanisms at lower energies. A significant effect on the yield ratio is also observed by turning off the hadronic cascade time, i.e., lower value of $t_{max}$. The non-monotonic energy dependence of the yield ratio $\mathcal{O}_{K^{\pm}-\Xi^{-}(\bar \Xi^{+})-\phi-\Lambda (\bar \Lambda)}$ of strange hadrons could be related to the large strange quark density fluctuations near the CEP or the first-order phase transition. The double yield ratios considered in the present study based on AMPT-SM model do not show any non-monotonic behaviors and thus providing a baseline for the search of CEP. The more realistic equation of state based dynamical modeling is still required for the heavy-ion collisions in order to extract the definite physics conclusion about the non-monotonic behavior.

\section{Conclusion}\label{sec5}

We reported the production of $K^{\pm}$, $\Xi^-(\bar \Xi)$, $\phi$ and $\Lambda (\bar \Lambda)$ in {\auau} collisions at {\sqrtsNN} = 7.7, 11.5, 14.5, 19.6, 27, 39, 54.4, 62.4, and 200 GeV from A Multi-Phase Transport model with string melting (AMPT-SM). We presented the transverse momentum ({\ppt}) spectra of these strange hadrons with a new set of parameters and by varying the hadronic cascade time ($t_{max}$) in the AMPT-SM model and found that the AMPT-SM model reasonably describe the STAR experimental data. No significant effect of $t_{max}$ is seen in the {\ppt} spectra except $\Xi^-$ and $\phi$ spectra at very low {\ppt} bin ($p_\mathrm{T} <$ 1 GeV/$c$). On the other hand, it is predicted that the yield ratios of strange quark, $\mathcal{O}_{K^{\pm}-\Xi^{-}(\bar \Xi^{+})-\phi-\Lambda (\bar \Lambda)}$ can be used to probe the strange quark density fluctuations in heavy-ion collisions. We analyzed the collision energy dependence of the strange hadron yield ratios $\mathcal{O}_{K^{\pm}-\Xi^{-}(\bar \Xi^{+})-\phi-\Lambda (\bar \Lambda)}$ in most central (0-10\%) and peripheral (40-60\%) {\auau} collisions from AMPT-SM model and compared with the experimental data. We found that the values of yield ratio $\mathcal{O}_{K^{\pm}-\Xi^{-}(\bar \Xi^{+})-\phi-\Lambda (\bar \Lambda)}$ from AMPT-SM model is much smaller than the experimental data and the model completely fails to describe even quantitatively the collision energy dependence of the ratio $\mathcal{O}_{K^{\pm}-\Xi^{-}(\bar \Xi^{+})-\phi-\Lambda (\bar \Lambda)}$. For a crossover transition between the Quark-Gluon Plasma (QGP) and hadronic matter, the double yield ratios considered in the present study based on AMPT-SM model do not show any non-monotonic behaviors and thus providing a baseline for the search of CEP, because there is no first-order or second-order phase transition in the AMPT model. The dynamical modeling based on more realistic equation of state is still required for the heavy-ion collisions in order to study the fluctuations and hence, extract the definite physics conclusion about the non-monotonic behavior.



\end{multicols}{}

\begin{thebibliography}{99}


\bibitem{BRAHMS:2004adc}
I.~Arsene \textit{et al.} [BRAHMS],
Nucl. Phys. A \textbf{757}, 1-27 (2005)

\bibitem{PHENIX:2004vcz}
K.~Adcox \textit{et al.} [PHENIX],
Nucl. Phys. A \textbf{757}, 184-283 (2005)

\bibitem{PHOBOS:2004zne}
B.~B.~Back \textit{et al.} [PHOBOS],
Nucl. Phys. A \textbf{757}, 28-101 (2005)


\bibitem{STAR:2005gfr}
J.~Adams \textit{et al.} [STAR],
Nucl. Phys. A \textbf{757}, 102-183 (2005)

\bibitem{Cleymans:1999st}
J.~Cleymans and K.~Redlich,
Phys. Rev. C \textbf{60}, 054908 (1999)

\bibitem{Becattini:2005xt}
F.~Becattini, J.~Manninen and M.~Gazdzicki,
Phys. Rev. C \textbf{73}, 044905 (2006)

\bibitem{Andronic:2005yp}
A.~Andronic, P.~Braun-Munzinger and J.~Stachel,
Nucl. Phys. A \textbf{772}, 167-199 (2006)



\bibitem{Shuryak:2014zxa}
E.~Shuryak,
Rev. Mod. Phys. \textbf{89}, 035001 (2017)



\bibitem{Braun-Munzinger:2015hba}
P.~Braun-Munzinger, V.~Koch, T.~Sch\"afer and J.~Stachel,
Phys. Rept. \textbf{621}, 76-126 (2016)


\bibitem{Fodor:2004nz}
Z.~Fodor and S.~D.~Katz,
JHEP \textbf{04}, 050 (2004)

\bibitem{Asakawa:1989bq}
M.~Asakawa and K.~Yazaki,
Nucl. Phys. A \textbf{504}, 668-684 (1989)

\bibitem{Stephanov:1998dy}
M.~A.~Stephanov, K.~Rajagopal and E.~V.~Shuryak,
Phys. Rev. Lett. \textbf{81}, 4816-4819 (1998)

\bibitem{Hatta:2002sj}
Y.~Hatta and T.~Ikeda,
Phys. Rev. D \textbf{67}, 014028 (2003)


\bibitem{Gavai:2008zr}
R.~V.~Gavai and S.~Gupta,
Phys. Rev. D \textbf{78}, 114503 (2008)

\bibitem{Asakawa:2008ti}
M.~Asakawa, S.~A.~Bass, B.~Muller and C.~Nonaka,
Phys. Rev. Lett. \textbf{101}, 122302 (2008)

\bibitem{Asakawa:2000wh}
M.~Asakawa, U.~W.~Heinz and B.~Muller,
Phys. Rev. Lett. \textbf{85}, 2072-2075 (2000)

\bibitem{Yu:2018kvh}
N.~Yu, D.~Zhang and X.~Luo,
Chin. Phys. C \textbf{44}, no.1, 014002 (2020)

\bibitem{Sun:2017xrx}
K.~J.~Sun, L.~W.~Chen, C.~M.~Ko and Z.~Xu,
Phys. Lett. B \textbf{774}, 103-107 (2017)

\bibitem{Sun:2018jhg}
K.~J.~Sun, L.~W.~Chen, C.~M.~Ko, J.~Pu and Z.~Xu,
Phys. Lett. B \textbf{781}, 499-504 (2018)

\bibitem{Shao:2019xpj}
T.~Shao, J.~Chen, C.~M.~Ko and K.~J.~Sun,
Phys. Lett. B \textbf{801}, 135177 (2020)





\bibitem{STAR:2009sxc}
B.~I.~Abelev \textit{et al.} [STAR],
Phys. Rev. C \textbf{81}, 024911 (2010)


\bibitem{STAR:2010vob}
M.~M.~Aggarwal \textit{et al.} [STAR],
[arXiv:1007.2613 [nucl-ex]].

\bibitem{STAR:2019bjj}
J.~Adam \textit{et al.} [STAR],
Phys. Rev. C \textbf{102}, no.3, 034909 (2020)

\bibitem{Ashraf:2016fjl}
M.~U.~Ashraf,
J. Phys. Conf. Ser. \textbf{668}, no.1, 012095 (2016)

\bibitem{Ashraf:2021nkb}
M.~U.~Ashraf [STAR],
Nucl. Phys. A \textbf{1005}, 121815 (2021)

\bibitem{STAR:2019vcp}
J.~Adam \textit{et al.} [STAR],
Phys. Rev. C \textbf{101}, no.2, 024905 (2020)

\bibitem{Rafelski:1982pu}
J.~Rafelski and B.~Muller,
Phys. Rev. Lett. \textbf{48} (1982), 1066
[erratum: Phys. Rev. Lett. \textbf{56} (1986), 2334]

\bibitem{Ahmad:1991nv}
S.~Ahmad, B.~E.~Bonner, C.~S.~Chan, J.~M.~Clement, S.~V.~Efremov, E.~Efstathiadis, S.~E.~Eiseman, A.~Etkin, K.~J.~Foley and R.~W.~Hackenburg, \textit{et al.}
Phys. Lett. B \textbf{382} (1996), 35-39
[erratum: Phys. Lett. B \textbf{386} (1996), 496-496]

\bibitem{Ahmad:1998sg}
S.~Ahmad, B.~E.~Bonner, S.~V.~Efremov, G.~S.~Mutchler, E.~D.~Platner and H.~W.~Themann,
Nucl. Phys. A \textbf{636} (1998), 507-524

\bibitem{E-802:1999ewk}
L.~Ahle \textit{et al.} [E-802 and E-866],
Phys. Rev. C \textbf{60} (1999), 044904

\bibitem{E917:2001eko}
B.~B.~Back \textit{et al.} [E917],
Phys. Rev. Lett. \textbf{87} (2001), 242301

\bibitem{Albergo:2002tn}
S.~Albergo, R.~Bellwied, M.~Bennett, D.~Boemi, B.~Bonner, H.~Caines, W.~Christie, S.~Costa, H.~J.~Crawford and M.~Cronqvist, \textit{et al.}
Phys. Rev. Lett. \textbf{88} (2002), 062301

\bibitem{E895:2003qcm}
P.~Chung \textit{et al.} [E895],
Phys. Rev. Lett. \textbf{91} (2003), 202301

\bibitem{NA49:2002pzu}
S.~V.~Afanasiev \textit{et al.} [NA49],
Phys. Rev. C \textbf{66} (2002), 054902

\bibitem{NA57:2004nxc}
F.~Antinori \textit{et al.} [NA57],
Phys. Lett. B \textbf{595} (2004), 68-74

\bibitem{STAR:2002fhx}
C.~Adler \textit{et al.} [STAR],
Phys. Rev. Lett. \textbf{89} (2002), 092301

\bibitem{PHENIX:2002svd}
K.~Adcox \textit{et al.} [PHENIX],
Phys. Rev. Lett. \textbf{89} (2002), 092302

\bibitem{ALICE:2013cdo}
B.~B.~Abelev \textit{et al.} [ALICE],
Phys. Rev. Lett. \textbf{111} (2013), 222301

\bibitem{Chen:2018tnh}
J.~Chen, D.~Keane, Y.~G.~Ma, A.~Tang and Z.~Xu,
Phys. Rept. \textbf{760} (2018), 1-39

\bibitem{Becattini:2000jw}
F.~Becattini, J.~Cleymans, A.~Keranen, E.~Suhonen and K.~Redlich,
Phys. Rev. C \textbf{64}, 024901 (2001)

\bibitem{Braun-Munzinger:2001alt}
P.~Braun-Munzinger, J.~Cleymans, H.~Oeschler and K.~Redlich,
Nucl. Phys. A \textbf{697}, 902-912 (2002)

\bibitem{Redlich:2001kb}
K.~Redlich and A.~Tounsi,
Eur. Phys. J. C \textbf{24}, 589-594 (2002)

\bibitem{Li:2016uvu}
F.~Li and C.~M.~Ko,
Phys. Rev. C \textbf{95}, no.5, 055203 (2017)

\bibitem{Steinheimer:2012gc}
J.~Steinheimer and J.~Randrup,
Phys. Rev. Lett. \textbf{109}, 212301 (2012)

\bibitem{Steinheimer:2013xxa}
J.~Steinheimer, J.~Randrup and V.~Koch,
Phys. Rev. C \textbf{89}, no.3, 034901 (2014)

\bibitem{Ko:2018ekv}
C.~M.~Ko,
EPJ Web Conf. \textbf{171}, 03002 (2018)

\bibitem{Sorge:1995dp}
H.~Sorge,
Phys. Rev. C \textbf{52}, 3291-3314 (1995)

\bibitem{Sorge:1997nv}
H.~Sorge,
Phys. Lett. B \textbf{402}, 251-256 (1997)

\bibitem{Bass:1998ca}
S.~A.~Bass, M.~Belkacem, M.~Bleicher, M.~Brandstetter, L.~Bravina, C.~Ernst, L.~Gerland, M.~Hofmann, S.~Hofmann and J.~Konopka, \textit{et al.}
Prog. Part. Nucl. Phys. \textbf{41}, 255-369 (1998)

\bibitem{Bleicher:1999xi}
M.~Bleicher, E.~Zabrodin, C.~Spieles, S.~A.~Bass, C.~Ernst, S.~Soff, L.~Bravina, M.~Belkacem, H.~Weber and H.~Stoecker, \textit{et al.}
J. Phys. G \textbf{25}, 1859-1896 (1999)

\bibitem{Kahana:1996ssp}
S.~H.~Kahana, D.~E.~Kahana, Y.~Pang and T.~J.~Schlagel,
Ann. Rev. Nucl. Part. Sci. \textbf{46}, 31-70 (1996)

\bibitem{Li:1997pu}
B.~A.~Li and C.~M.~Ko,
Nucl. Phys. A \textbf{630}, 556-562 (1998)

\bibitem{Nara:2019crj}
Y.~Nara,
EPJ Web Conf. \textbf{208}, 11004 (2019)

\bibitem{Lin:2004en}
Z.~W.~Lin, C.~M.~Ko, B.~A.~Li, B.~Zhang and S.~Pal,
Phys. Rev. C \textbf{72}, 064901 (2005)

\bibitem{Lin:2001zk}
Z.~w.~Lin and C.~M.~Ko,
Phys. Rev. C \textbf{65}, 034904 (2002)

\bibitem{Chen:2004vha}
L.~W.~Chen, V.~Greco, C.~M.~Ko and P.~F.~Kolb,
Phys. Lett. B \textbf{605}, 95-100 (2005)

\bibitem{Wang:1991hta}
X.~N.~Wang and M.~Gyulassy,
Phys. Rev. D \textbf{44}, 3501-3516 (1991)


\bibitem{Andersson:1983jt}
B.~Andersson, G.~Gustafson and B.~Soderberg,
Z. Phys. C \textbf{20} (1983), 317

\bibitem{Andersson:1983ia}
B.~Andersson, G.~Gustafson, G.~Ingelman and T.~Sjostrand,
Phys. Rept. \textbf{97} (1983), 31-145

\bibitem{Zhang:1999bd}
B.~Zhang, C.~M.~Ko, B.~A.~Li and Z.~w.~Lin,
Phys. Rev. C \textbf{61} (2000), 067901

\bibitem{Lin:2000cx}
Z.~w.~Lin, S.~Pal, C.~M.~Ko, B.~A.~Li and B.~Zhang,
Phys. Rev. C \textbf{64} (2001), 011902


\bibitem{Lin:2021mdn}
Z.~W.~Lin and L.~Zheng,
Nucl. Sci. Tech. \textbf{32} (2021) no.10, 113


\bibitem{Shao:2020sqr}
T.~Shao, J.~Chen, C.~M.~Ko and Z.~W.~Lin,
Phys. Rev. C \textbf{102} (2020) no.1, 014906

\bibitem{Ashraf:2022rjt}
M.~U.~Ashraf, J.~Tariq, S.~Ikram, A.~M.~Khan, J.~Butt and S.~Zain,
[arXiv:2211.14795 [hep-ph]].

\bibitem{He:2017tla}
Y.~He and Z.~W.~Lin,
Phys. Rev. C \textbf{96} (2017) no.1, 014910

\bibitem{Xu:2012gf}
J.~Xu, L.~W.~Chen, C.~M.~Ko and Z.~W.~Lin,
Phys. Rev. C \textbf{85} (2012), 041901

\bibitem{Pradhan:2021zbt}
G.~S.~Pradhan, R.~Rath, R.~Scaria and R.~Sahoo,
Phys. Rev. C \textbf{105}, no.5, 054905 (2022)

\bibitem{Sun:2017ooe}
K.~J.~Sun and L.~W.~Chen,
Phys. Rev. C \textbf{95}, no.4, 044905 (2017)

\bibitem{STAR:2017sal}
L.~Adamczyk \textit{et al.} [STAR],
Phys. Rev. C \textbf{96} (2017) no.4, 044904

\bibitem{STAR:2015vvs}
L.~Adamczyk \textit{et al.} [STAR],
Phys. Rev. C \textbf{93} (2016) no.2, 021903

\bibitem{STAR:2006egk}
J.~Adams \textit{et al.} [STAR],
Phys. Rev. Lett. \textbf{98} (2007), 062301

\bibitem{STAR:2007mum}
B.~I.~Abelev \textit{et al.} [STAR],
Phys. Rev. Lett. \textbf{99} (2007), 112301

\bibitem{STAR:2002hpr}
C.~Adler \textit{et al.} [STAR],
Phys. Lett. B \textbf{595} (2004), 143-150

\bibitem{STAR:2008med}
B.~I.~Abelev \textit{et al.} [STAR],
Phys. Rev. C \textbf{79} (2009), 034909

\bibitem{Wheaton:2004qb}
S.~Wheaton and J.~Cleymans,
Comput. Phys. Commun. \textbf{180} (2009), 84-106

\bibitem{NA49:2007stj}
C.~Alt \textit{et al.} [NA49],
Phys. Rev. C \textbf{77} (2008), 024903

\bibitem{NA49:2008goy}
C.~Alt \textit{et al.} [NA49],
Phys. Rev. C \textbf{78} (2008), 044907

\bibitem{NA49:2008ysv}
C.~Alt \textit{et al.} [NA49],
Phys. Rev. C \textbf{78} (2008), 034918

\end{thebibliography}
\end{document}